\documentclass[epj]{svjour}
\usepackage{graphics}
\usepackage{graphicx,t1enc}
\usepackage{epsfig}
\usepackage{amssymb,amsmath}

\begin{document}
\title{Interplay of noise and coupling in heterogeneous ensembles of phase oscillators}
\author{Dami\'an H. Zanette}
\institute{Consejo Nacional de Investigaciones Cient\'{\i}ficas y
T\'ecnicas, Centro At\'omico Bariloche and Instituto Balseiro, 8400
Bariloche, R\'{\i}o Negro, Argentina}
\date{Received: date / Revised version: date}

\authorrunning{D. H. Zanette}
\titlerunning{Noise and  coupling in heterogeneous oscillator
ensembles}

\abstract{We study the effects of noise on the collective dynamics
of an ensemble of coupled phase oscillators whose natural
frequencies are all identical, but whose coupling strengths are not
the same all over the ensemble. The intensity of noise can also be
heterogeneous, representing diversity in the individual responses to
external fluctuations. We show that the desynchronization transition
induced by noise may be completely suppressed, even for arbitrarily
large noise intensities, is the distribution of coupling strengths
decays slowly enough for large couplings. Equivalently, if the
response to noise of a sufficiently large fraction of the ensemble
is weak enough, desynchronization cannot occur. The two effects
combine with each other when the response to noise and the coupling
strength of each oscillator are correlated. This combination is
quantitatively characterized and illustrated with explicit examples.
} \maketitle

The statistical description of a large physical system, formed by a
multitude of interacting elements, is admittedly simpler if all the
elements are mutually identical. In such a homogeneous ensemble,
every element is representative of any other, which simplifies the
calculation of average quantities and collective properties, both
static and dynamic. Many applications require however the
consideration of heterogeneous ensembles to account for various
sources of diversity, from dispersion in the parameters that govern
the individual dynamics of the elements and their interaction, to
differences in the external influences, such as noise, that affect
each element independently \cite{fromcells}.

Kuramoto's theory for the synchronization of coupled oscillators
\cite{kura} is an instance where heterogeneity plays a key role in
the collective behaviour of a large ensemble. The synchronization
transition, between states of incoherent and coherent dynamics,
results from the competition of the strength of coupling and the
dispersion in the natural frequencies of individual oscillators
\cite{win}. Further sources of heterogeneity are, for instance, the
topology of connection patterns \cite{red1,red2} and diversity in
the interaction laws between oscillator pairs \cite{fases}.

In this paper, we study an extension of Kuramoto's model, with two
additional sources of diversity, which still admits a fully
analytical treatment. Consider an ensemble of $N$ phase oscillators,
each of them characterized by a phase $\phi_i \in [0,2\pi)$. The
dynamics of phases is given by
\begin{equation} \label{eq1}
\dot \phi_i = \omega_i + \frac{k_i}{N} \sum_{j=1}^N \sin
(\phi_j-\phi_i) +\xi_i(t) ,
\end{equation}
$i=1,\dots, N$, where $\omega_i$ is the natural frequency of
oscillator $i$, $k_i>0$ is the coupling strength for the same
oscillator. In the standard Kuramoto's model, all oscillators have
identical coupling strengths, $k_i=K$ for all $i$. The
non-correlated  Gaussian noises $\xi_i(t)$ have zero mean, and
$\langle \xi_i(t) \xi_j (t') \rangle =2s_i \delta_{ij} \delta
(t-t')$. Note that here we admit that the intensity of noise,
$s_i>0$, may be different for each oscillator, standing for
diversity in the response to external fluctuations.

As in the standard  model \cite{kura,nos}, equation (\ref{eq1}) can
be recast as
\begin{equation} \label{evol}
\dot \phi_i = \omega_i +  k_i  \sigma \sin(\Phi -\phi_i)+\xi_i(t),
\end{equation}
with
\begin{equation} \label{sigma}
\sigma \exp({\rm i} \Phi ) = \frac{1}{N} \sum_{j=1}^N  \exp({\rm i}
\phi_j ).
\end{equation}
The non-negative number $\sigma$ is the Kuramoto order parameter,
which characterizes the synchronization transition: $\sigma=0$
represents an incoherent state with uniform distribution of phases,
while $\sigma>0$ reveals a certain degree of organization in the
phases, that we associate with synchronization.

In previous work \cite{pais1,pais2}, we have studied the extended
model (\ref{evol}) in the absence of noise, $s_i\equiv 0$ for all
$i$. Heterogeneous coupling strengths make it possible that an
oscillator whose natural frequency is far from the synchronization
frequency becomes nevertheless entrained in synchronized behaviour
if its coupling strength is large enough. Synchronization is thus
enhanced. If, on the other hand, oscillators close to the
synchronization frequency have systematically low coupling
strengths, synchronization may be completely suppressed.

Here, we focus our attention on the case where the natural
frequencies of all oscillators are identical, $\omega_i=\omega$ for
all $i$. The transformation $\phi_i \to \phi_i+\omega t$ for all $i$
makes it possible to take, without generality loss, $\omega\equiv
0$. Now, moreover, external noise is present: $s_i \neq 0$.
Heterogeneity in the intrinsic dynamics of the oscillators, given by
their natural frequencies, is thus replaced by fluctuations in the
form of additive noise.

In order to provide a statistical description of our system in the
limit $N\to \infty$, we introduce $n(k,s ;\phi,t)$, the density of
oscillators with coupling strength $k$ and noise intensity $s$
which, at time $t$, have phase $\phi$. Standard results from the
theory of stochastic processes \cite{stoch} establish that, if
equation (\ref{eq1}) governs the dynamics of phases, the density
$n(k,s;\phi,t)$ satisfies the Fokker-Planck equation
\begin{equation}
\partial_t n = s \partial^2_{\phi }  n+k\sigma \partial_\phi
[n \sin (\phi-\Phi) ] .
\end{equation}
The stationary solution to this equation, obtained from $\partial_t
n \equiv 0$, reads
\begin{equation} \label{nst}
n(k,s;\phi) = \frac{1}{2\pi I_0(k\sigma /s)} \exp\left[
\frac{k\sigma}{s} \cos (\phi-\Phi)\right],
\end{equation}
where $I_\nu (x)$ is the order-$\nu$ modified Bessel function of the
first kind \cite{abramo}. This solution, which is expected to
represent the long-time, equilibrium distribution of phases, remains
however a formal expression, since both $\sigma$ and $\Phi$ are
still unknown.

To obtain  self-consistent values for $\sigma$ and $\Phi$, we
transform equation (\ref{sigma}) into its continuous version by
using the oscillator density $n$. To do this, we assume that the
interaction strengths $k_i$ and the noise intensities $s_i$ are
assigned over the ensemble following a prescribed distribution
$W(k,s)$.  The normalization of this distribution requires
\begin{equation} \label{norm}
\int_0^\infty  d  k \int_0^\infty   d  s \, W(k,s)  =1.
\end{equation}
Note that, since both the interaction strength and the noise
intensity of a given oscillator measure its response to extrinsic
actions --respectively, the rest of the ensemble and fluctuations--
it is not unlikely that they are correlated attributes, so that
$W(k,s)$ cannot generally be factorized. The sum over oscillators in
equation (\ref{sigma}) becomes thus a multiple integral over  $k$,
$s$, and the phase $\phi$. Using the equilibrium density of equation
(\ref{nst}), we get
\begin{eqnarray} \label{long}
\sigma = &\displaystyle{\int_0^{2\pi}  d  \phi \int_0^\infty  d  k
\int_0^\infty  d  s  \, \frac{W(k,s)}{2\pi I_0(k\sigma /s)}}\nonumber  \\ \nonumber  \\
& \times \exp\left(  \displaystyle{\frac{k\sigma}{s}} \cos
\phi\right) \exp\left({\rm i} \phi \right).
\end{eqnarray}

Due to the fact that the integrand is $2\pi$-periodic in $\phi$ and
that the integral over the phase runs over a whole period, the value
of the collective phase $\Phi$ is arbitrary; we have chosen
$\Phi=0$. Performing the integral over $\phi$, we find
\begin{equation} \label{sig}
\sigma = \int_0^\infty  d  k \int_0^\infty  d  s \, W(k,s)
\frac{I_1(k\sigma /s)}{ I_0(k\sigma /s)}.
\end{equation}
This implicit equation for $\sigma$ is our main result. It
determines the order parameter in terms of the density of
interaction strengths and noise intensities, $W(k,s)$. It always has
a trivial solution $\sigma=0$ which, according to equation
(\ref{nst}), corresponds to a uniform phase distribution
$n(k,s;\phi)= (2\pi)^{-1}$. Synchronized states are those where, on
the contrary, $\sigma \neq 0$.

For homogeneous interaction strengths and noise intensities, $k_i=K$
and $s_i=S$ for all $i$, we have $W(k,s)= \delta(k-K) \delta(s-S)$.
Equation (\ref{sig}) reduces to
\begin{equation}
\sigma = I_1(K\sigma /S) / I_0(K\sigma /S),
\end{equation}
a limit that has already been analyzed in the literature
\cite{fromcells}. For large intensity noises, the only solution is
$\sigma=0$, and the oscillator ensemble is unsynchronized. On the
other hand, a positive solution exists if $S$ is sufficiently small.
This solution appears through a pitchfork bifurcation at the
critical noise intensity $S_c=K/2$. It approaches $\sigma=1$,
corresponding to complete synchronization, as $S\to 0$.

To study how this scenario changes when interaction strengths and
noise intensities are not the same all over the ensemble, we first
analyze the behaviour of the right-hand side of equation (\ref{sig})
--which, for conciseness, we call $J(\sigma)$-- as a function of
$\sigma$. We assume that the distribution $W(k,s)$, which must
satisfy the normalization (\ref{norm}), is regular enough as to
warrant the conclusions drawn in the following. Since $I_0(0)=1$ and
$I_1(0)=0$, $J(\sigma)$ vanishes as $\sigma\to 0$. The ratio
$I_1(x)/I_0(x)$, in turn, tends to one as $x\to \infty$. Therefore,
by virtue of equation (\ref{norm}), $J(\sigma)$ also approaches one
as $\sigma \to \infty$. Assuming that $J(\sigma)$ varies
monotonically from zero to one as $\sigma$ grows from zero to
infinity, equation (\ref{sig}) will have a single solution at
$\sigma=0$ if the slope of $J(\sigma)$ at that point is lower than
one. On the other hand, a non-trivial solution will exist if the
slope is greater than one. Straightforward calculation of the
derivatives of the Bessel functions at zero shows that this
condition, which fixes the threshold for synchronization, is
equivalent to
\begin{equation} \label{crit}
1 = \int_0^\infty  d  k \int_0^\infty  d  s \, \frac{k}{2s} W(k,s) .
\end{equation}
This  equation must be interpreted as a condition to be fulfilled by
the parameters that define the distribution $W(k,s)$. In  parameter
space, it determines the boundary between regions of synchronized
and unsynchronized dynamics, namely, the desynchronization boundary.

To appraise how equation (\ref{crit}) works, let us analyze two
extreme situations. In the first situation, the coupling strength
and the noise intensity  of each oscillator are fully uncorrelated
attributes, so that their distribution can be factorized as $W(k,s)
= W_1(k) W_2(s)$. Suppose also that $W_2(s) = \delta (s-S) $, so
that the noise intensity is equal to $S$ all over the ensemble. In
this case, equation (\ref{crit}) establishes that the critical noise
intensity is
\begin{equation} \label{Ac}
S_c = \frac{1}{2} \int_0^\infty  d  k  \, k W_1(k) .
\end{equation}
From this result we draw our first important conclusion: in an
ensemble with heterogeneous coupling, noise is able to suppress
synchronization as long as the function $k W_1(k)$ is integrable
over $(0,\infty)$. In other words, if the distribution of coupling
strengths decays slowly enough for $k\to \infty$ --namely, as $W_1
\sim k^{-p}$ with $1<p<2$-- synchronized behaviour persists even for
arbitrarily large external fluctuations.

For more general forms of $W_2(s)$, the factorization of $W(k,s)$
implies that the right-hand side of equation (\ref{crit}) is a
product of two integrals, respectively over $k$ and $s$. The
equation can hold only if the two integrals converge. In particular,
the function $s^{-1} W_2 (s)$ must be integrable over $(0,\infty)$.
Now, therefore, fluctuations are able to suppress synchronized
behaviour only if $W_2(s)$ vanishes for $s\to 0$. Specifically, if
the fraction of the ensemble with noise intensities below a small
threshold $\delta s$ is proportional to $\delta s$ (or larger), too
many oscillators are subject to too weak noise, and fluctuations
cannot inhibit synchronization.

At the opposite extreme, we examine the case where the correlation
between coupling strength and noise intensity is so strong that one
of the two attributes is a given function of the other. We take
$s\equiv \zeta (k)$, so that $W(k,s)= W_1 (k) \delta [s - \zeta
(k)]$. Equation (\ref{crit}) becomes
\begin{equation} \label{correl}
1 =  \int_0^\infty  d  k  \, \frac{k}{2 \zeta (k) } W_1(k) .
\end{equation}
A necessary condition for this equation to hold is now that the
function $k W_1(k)/\zeta (k)$ is integrable over $(0,\infty)$. In
particular, the distribution  of coupling strengths may decay as
slowly as to make the product $kW_1(k)$ non-integrable. But if, at
the same time, the intensity of noise grows sufficiently fast with
the coupling strength, the desynchronization transition can still
take place. As for the integrability condition at $k \to 0$, the
integral may diverge if the noise intensity $\zeta (k)$ exhibits a
sufficiently fast decay with the coupling strength.

Let us illustrate these conclusions with results for the order
parameter $\sigma$ corresponding to some specific forms of the
distribution of coupling strengths and noise intensities, chosen in
such a way as to exemplify the different situations analyzed above.
Generally, equation (\ref{sig}) must be solved by numerical means,
as explicit expressions for the involved integrals are usually not
known.

\begin{figure}\includegraphics[width=\columnwidth]{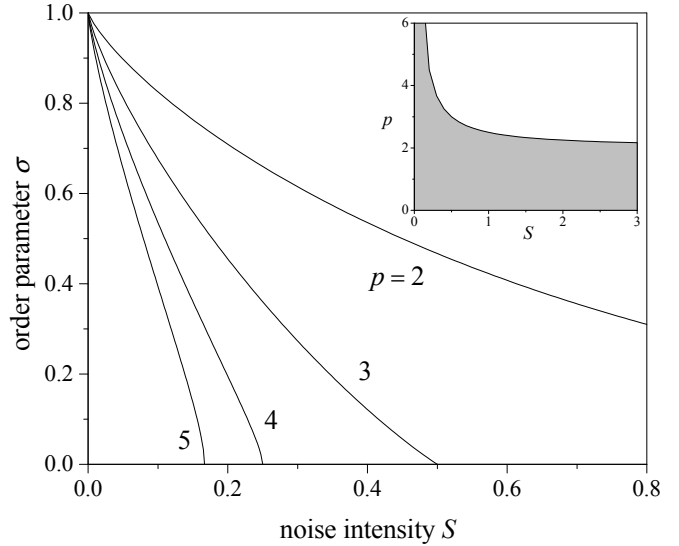}
\caption{Kuramoto order parameter
$\sigma$ as a function of the noise intensity $S$ in an ensemble of
coupled oscillators with coupling strength distribution $W_1(k) =
(p-1) (1+k)^{-p}$, for several values of $p$. The inset shows the
synchronization region ($\sigma \neq 0$; shaded) in the parameter
plane $(S,p)$.} \label{fig1}
\end{figure}

Consider first the case where the noise intensity $S$ is the same
all over the ensemble. For the distribution of coupling strengths,
we take
\begin{equation}  \label{13}
W_1(k) = (p-1) (1+k)^{-p}
\end{equation}
with $p>1$, i.e. a power-law decaying function of $k$. Figure
\ref{fig1} shows the order parameter as a function of the noise
intensity for several values of $p$. As expected, in all cases, the
degree of synchronization decreases with noise. The critical noise
intensity at which the order parameter vanishes,
$S_c=[2(p-2)]^{-1}$, is well defined for $p>2$. Note also that the
behaviour of $\sigma$ at the critical point varies with $p$.
Approximating equation (\ref{sig}) for $\sigma \approx 0$, in fact,
we find $ \sigma \approx \alpha (S_c-S)^{1/(p-2)}$, with $\alpha$ a
constant. For $1< p\le 2$, the order parameter decays indefinitely,
never reaching zero, as $S$ grows. The inset of Figure \ref{fig1}
shows, in gray, the zone of parameter space where synchronized
dynamics occurs. This phase diagram suggests that the behaviour of
the order parameter as a function of $p$, for fixed noise intensity,
would be qualitatively the same as shown in the main plot as a
function of $S$.

\begin{figure}
\includegraphics[width=\columnwidth]{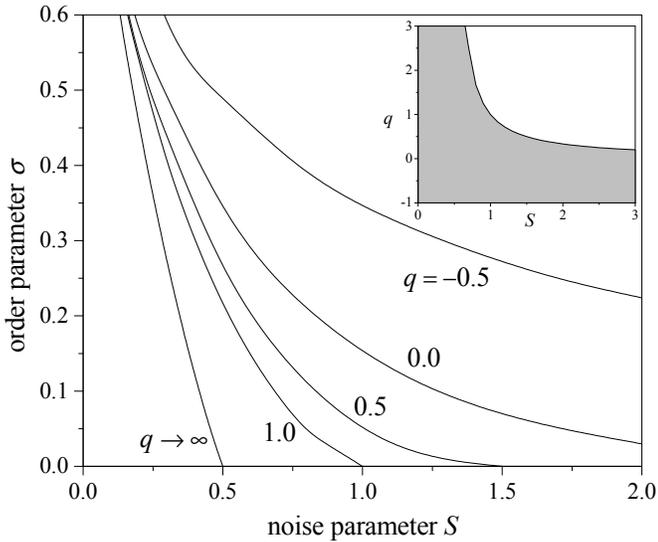} \caption{Kuramoto order
parameter $\sigma$ as a function of the parameter $S$ in an ensemble
of coupled oscillators with noise intensity distribution
$W_2(s)\propto s^q \exp[-(q+1)s/S]$, for several values of $q$. The
coupling strength distribution is as in fig. \ref{fig1}, with $p=3$.
The inset shows the synchronization region ($\sigma \neq 0$; shaded)
in the parameter plane $(S,q)$.} \label{fig2}
\end{figure}

Next, we consider that the noise intensity is heterogeneous, but
still uncorrelated to the coupling strength. We take
\begin{equation}
W_2(s)= w_0 s^q \exp[-(q+1)s/S],
\end{equation}
where $w_0$ is a normalization constant. The exponent $q>-1$,
controls the shape of the distribution  at $s=0$. For $q>0$, the
distribution vanishes at the origin and has a maximum at $s=S
q/(q+1)$. As $q\to \infty$, it approaches $W_2(s)=\delta (s-S)$, the
case considered in the preceding paragraph. For $q=0$, the
distribution is a purely decaying exponential, and for $q<0$ it
diverges as $s$ approaches zero. As for the distribution of coupling
strengths, we take the same as in equation (\ref{13}) with $p=3$,
namely, $W_1(k) = 2 (1+k)^{-3}$. In Figure \ref{fig2}, we plot the
order parameter $\sigma$ as a function of the noise parameter $S$,
for several values of the exponent $q$. The curve for $q\to \infty$
coincides with that of Figure \ref{fig1} for $p=3$. The critical
noise parameter at which $\sigma$ vanishes, $S_c$, shifts to higher
values as $q$ decreases. Its analytical evaluation, in fact, shows
that the desynchronization transition takes place at $S_c =
(1+q)/2q$. As expected, $q=0$ is the largest value of $q$ for which
noise is not able to suppress synchronization. For smaller
exponents, the distribution of noise intensities does not vanish at
$s=0$, and synchronized behaviour persists even for arbitrarily
large values of $S$.

\begin{figure}
\includegraphics[width=\columnwidth]{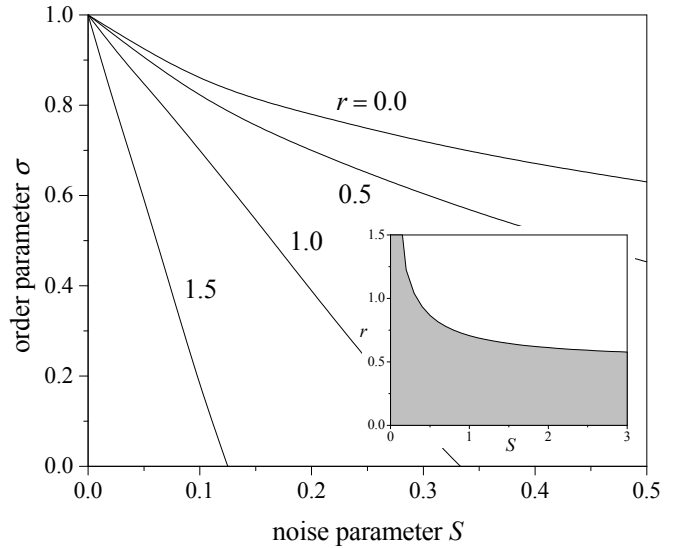} \caption{Kuramoto order
parameter $\sigma$ as a function of the parameter $S$ in an ensemble
of coupled oscillators with the distribution of coupling strengths
and noise intensities given in equation (\ref{zeta}), taking
$\zeta(k)= S (1+k)^r$, as a function of  $S$ and for several values
of $r$. The inset shows  the synchronization region ($\sigma \neq
0$; shaded) in the parameter plane $(S,r)$.} \label{fig3}
\end{figure}

Finally, we study a case where there is correlation between the
coupling strength and the noise intensity of each oscillator.
However, instead of taking --as in the derivation of equation
(\ref{correl})-- a deterministic relation between $k$ and $s$, we
fix
\begin{equation} \label{zeta}
W(k,s)=  \frac{s}{2\zeta^2(k)}  \exp[-s/\zeta (k)](1+k)^{-3/2} ,
\end{equation}
which combines the above form of $W_1(k)$, equation (\ref{13}), for
$p=3/2$ with an exponential function correlating $k$ and $s$. For a
given value of $k$, this function has a maximum at $s=\zeta(k)$; we
take $\zeta(k) \equiv S (1+k)^r$. The exponent $r$ controls how the
most frequent noise intensity $\zeta$ depends on the coupling
strength. For $r=0$, the correlation between $k$ and $s$ disappears.
The distribution of noise intensities becomes independent of $k$
and, because of the slow decay in the distribution of coupling
strengths, noise is not able to suppress synchronization. As $r$
grows positive, on the other hand, oscillators with larger coupling
strengths suffer, on the average, larger noise intensities, and
synchronization may be inhibited by noise. Analytical calculations
on equation (\ref{crit}) show that the desynchronization transition
takes place if $r>1/2$. In this situation, the critical value for
the noise parameter is $S_c=(4r^2-1)^{-1}$. Figure \ref{fig3}
displays the Kuramoto order parameter as a function of $S$, for
several values of the exponent $r$.

Our main conclusions can be summarized as follows. We have
considered a heterogeneous ensemble of coupled phase oscillators
subject to external fluctuations, where the coupling strength and
the effect of noise can be different for each oscillator. In
ensembles with homogeneous coupling synchronized behaviour emerges
spontaneously, but sufficiently large homogeneous fluctuations
inhibit synchronization \cite{fromcells}. We have shown here that,
on the other hand, homogeneous noise may not be able to inhibit
synchronization if the coupling strength is not the same for all
oscillators. Specifically, if the distribution of coupling strengths
decays  slowly enough for large couplings, synchronization persists
even under arbitrarily large fluctuations. A similar, complementary
effect takes place when noise intensities are in turn heterogeneous
over the  ensemble. An excess of oscillators with very small
response to noise can suppress unsynchronized behaviour, even when
the distribution of coupling strengths would allow for the
desynchronization transition under large homogeneous noise. In the
more generic situation where the coupling strength and the noise
intensity of each oscillator are correlated, the two attributes may
``control'' each other. Large couplings, which favor
synchronization, compete with large fluctuations, which tend to
inhibit coherent behaviour. The precise form of their correlation
defines whether the desynchronization transition exists or not.

Similar results were implicit in the analysis of oscillator
ensembles where both natural frequencies and coupling strengths are
heterogeneous, in the absence of noise \cite{pais1,pais2}. For
instance, if the distribution of coupling strengths at the
synchronization frequency decays slowly enough, the
desynchronization transition induced by a sufficiently flat
distribution of natural frequencies is suppressed, and synchronized
dynamics persists for arbitrarily flat distributions. This provides
a further example of the equivalent roles of diversity --in our
case, heterogeneous natural frequencies-- and noise, in the
collective dynamics of large ensembles of interacting elements
\cite{fromcells,tess1,tess2}.

\section*{Acknowledgements}
This work was partially supported by grants CONICET-PIP5115 and
ANPCyT-PICT2004-4-943, Argentina.

\end{document}